\documentclass[%
 reprint,
 amsmath,amssymb,
 aps,
]{revtex4-2}
\usepackage[thinc]{esdiff}
\usepackage{mathtools}
\usepackage{graphicx}% Include figure files
\usepackage{dcolumn}% Align table columns on decimal point
\usepackage{bm}% bold math
\usepackage{subfigure}

 \usepackage{wrapfig}
  \usepackage{comment}

\begin{document}

\preprint{APS/123-QED}

\title{What Does Nature Minimize In Every Incompressible Flow?}% Force line breaks with \\

\author{Haithem E. Taha}%
 %\email{hetaha@uci.edu}
 \author{Cody Gonzalez}
\affiliation{%
 University of California, Irvine\\
}%
%\collaboration{NASA (MUREP)}%\noaffiliation


\begin{abstract}
In this paper, we discover the fundamental quantity that Nature minimizes in almost all flows encountered in everyday life: river, rain, flow in a pipe, blood flow, airflow over an airplane, etc. We show that the norm of the pressure gradient over the field is minimum at every instant of time! We call it the principle of minimum pressure gradient (PMPG). The principle is deeply rooted in classical mechanics via \textit{Gauss’ principle of least constraint}. Therefore, while we prove mathematically that Navier-Stokes’ equation represents the necessary condition for minimization of the pressure gradient, the PMPG stands on its own philosophy independent of Navier-Stokes’. It turns any fluid mechanics problem into a minimization one. We demonstrate this intriguing property by solving three of the classical problems in fluid mechanics using the PMPG without resorting to Navier-Stokes’ equation. In fact, the inviscid version of the PMPG allowed solving the long-standing problem of the aerohydrodynamic lift over smooth cylindrical shapes where Euler's equation fails to provide a unique answer. Moreover, the result challenges the accepted wisdom about lift generation on an airfoil, which has prevailed over a century. The PMPG is expected to be transformative for theoretical modeling of fluid mechanics as it encodes a complicated nonlinear partial differential equation into a simple minimization problem. The principle even transcends Navier-Stokes’ equations in its applicability to non-Newtonian fluids with arbitrary constitutive relations and fluids subject to arbitrary forcing (e.g. electric or magnetic).
\end{abstract}

\maketitle

%\tableofcontents
\vspace{-0.2in}
\section{Introduction}
\vspace{-0.15in}
Remarkably, since the development of Navier-Stokes' equation (almost two centuries ago), there has not hitherto been a minimization principle for it that reveals the fundamental quantity that Nature minimizes in fluid motion. The reason for this long-standing gap is that the dominant variational principle in physics is Hamilton's principle of least action, which does not directly allow for non-conservative forces (polygenic forces that do not come from a scalar work function). Therefore, most of the variational principles of fluid mechanics in the literature were developed for ideal fluids (inviscid fluids governed by Euler's equations); e.g.,  \citep{Variational_Principles_Fluids1,Variational_Principles_Fluids2,Variational_Principles_Fluids_Hamilton_PoF,Seliger_Variational_Principles,Variational_Principles_Fluids5,Hamiltonian_Fluid_Mechanics,Variational_Principles_Morrison_Review}, which ignores important features (e.g., viscosity, turbulence, and other irreversible phenomena). 

There have been several efforts aiming at extending these variational formulations to account for dissipative/viscous forces \cite{Variational_Principles_Fluids_Stochastic,Variational_Principles_NS,Variational_Principles_Fluids_Stochastic_Gomes,Variational_Principles_Fluids_Stochastic2,Variational_Principles_NS_Nonholonomic,Variational_Principles_NS_2n,Variational_Principles_NS_Nonholonomic2}. However, these extensions do not directly follow from first principles; some ingenious mathematical manipulations are required to show the connection with the governing equations. So, often these variational formulations are imbued with a sense of \textit{ad hoc} and contrived treatments, which detracts from the beauty of analytical and variational formulations. It may be prudent to recall Salmon's statement: ``\textit{the existence of a Hamiltonian structure is, by itself, meaningless because any set of evolution equations can be written in canonical form}" by adding artificial variables. So, we view the few existing variational formulations of Navier-Stokes (e.g., \cite{Variational_Principles_Fluids_Stochastic,Variational_Principles_NS,Variational_Principles_Fluids_Stochastic_Gomes,Variational_Principles_Fluids_Stochastic2,Variational_Principles_NS_Nonholonomic,Variational_Principles_NS_2n,Variational_Principles_NS_Nonholonomic2}) as aesthetic mathematical constructions that managed to recover the (already known) governing equations, but may not provide new insights on the physics of fluids. In particular, it is not clear at all how one can use any of these variational formulations to solve (even simple) fluid mechanics problems (or inferring new concepts about the physics of fluids) without invoking the governing equations.

So, the main focus here is not on the mere development of a variational principle. Rather, we aim to \textit{discover} the fundamental quantity that Nature minimizes in every incompressible flow problem. The result is a variational principle of Navier-Stokes' equations that naturally stems from one of the first principles of mechanics: \textit{Gauss' principle of least constraint}. Relying on the fact that the pressure gradient force in Navier-Stokes' equations of motion is a \textit{constraint force} (i.e., whose sole role is to maintain the continuity constraint), we prove that the magnitude of the pressure gradient is minimum at every instant! We call it \textit{The Principle of Minimum Pressure Gradient} (PMPG).

In contrast to the previously devised variational principles of Navier-Stokes, (e.g., \cite{Variational_Principles_Fluids_Stochastic,Variational_Principles_NS,Variational_Principles_Fluids_Stochastic_Gomes,Variational_Principles_Fluids_Stochastic2,Variational_Principles_NS_Nonholonomic,Variational_Principles_NS_2n,Variational_Principles_NS_Nonholonomic2}), the PMPG is a true minimization principle (not just stationary). Hence, it reveals the fundamental quantity that Nature minimizes in every incompressible flow. Moreover, since the PMPG is philosophically independent of Navier-Stokes (in contrast to the previous variational principles that were mainly developed to recover Navier-Stokes' equation), it is more generic than the Navier-Stokes Newtonian framework: If a new constitutive model or an additional forcing are introduced, the Navier-Stokes equation must be modified whereas the PMPG remains valid with the exact same statement: the norm of the pressure gradient over the field is minimum at every instant.

The PMPG is expected to revolutionize fluid mechanics by turning any fluid mechanics problem into an optimization one where fluid mechanicians need not to apply Navier-Stokes' equations anymore. Rather, they merely need to minimize the objective/cost function. Moreover, because of its philosophical independence from the Newtonian formulation, the PMPG allows solving problems that are classically indeterminate using the current canonical formulations.  For example, in Sec. IV. C, we show that the inviscid version of the PMPG allows determination of aerohydrodynamic lift over smooth cylindrical shapes where Euler equation fails to provide a unique answer, solving an elusive problem that challenged fluid mechanicians over a century. 

\vspace{-0.2in}
\section{Gauss' Principle of Least Constraint}
\vspace{-0.2in}
Surprisingly, there are few published materials on Gauss' principle; Papastavridis wrote: ``\textit{In most of the $20^{\rm{th}}$ century English literature, GP [Gauss Principle] has been barely tolerated as a clever but essentially useless academic curiosity, when it was mentioned at all}" \cite{Papastavridis2014}. And only few efforts have adopted it in the $21^{\rm{st}}$ century \cite{Gauss_21century,Udwadia_Kalaba_Book}.

Consider the dynamics of $N$ constrained particles, each of mass $m_i$, such that we have a total of $n$ generalized coordinates (degrees of freedom) $\bm{q}$. The dynamics of these $N$ particles are governed by Newton's equations
\begin{equation}\label{eq:Newton}:
  m_i \bm{a}_i(\ddot{\bm{q}},\dot{\bm{q}},\bm{q}) = \bm{F}_i + \bm{R}_i \; \forall i\in\{1,..,N\},
\end{equation}
where $\bm{a}_i$ is the inertial acceleration of the $i^{\rm{th}}$ particle. The right hand side of the equation represents the total force acting on the particle, which is typically decomposed in analytical mechanics into:  (i) impressed forces $\bm{F}_i$, which are the directly applied (driving) forces (e.g., gravity, elastic, viscous); and (ii) constraint forces $\bm{R_i}$ whose raison d'etre is to enforce kinematical/geometrical constraints; they are passive or workless forces \cite{Lanczos_Variational_Mechanics_Book}. That is, they do not contribute to the motion abiding by the constraint; their main mission is to preserve the constraint (i.e., prevent any deviation from it). Examples include the force in a pendulum rod and the normal force acting on a particle sliding over a surface. 

Inspired by his method of least squares, Gauss asserted that the quantity
\begin{equation}\label{eq:Gauss}
  Z = \frac{1}{2} \sum_{i=1}^N m_i\left( \bm{a}_i- \frac{\bm{F}_i}{m_i} \right)^2
\end{equation}
is minimum with respect to the generalized accelerations $\ddot{\bm{q}}$ \cite[][pp. 911-912]{Papastavridis2014}. Note that the quantity $Z$ is nothing but $Z = \frac{1}{2} \sum_{i=1}^N \frac{\bm{R}_i^2}{m_i}$, i.e., the magnitude of constraint forces must be minimum---hence the name least constraint.

Gauss' principle is adroitly intuitive. In the absence of constraints, a particle follows the applied acceleration/force $\frac{\bm{F}}{m}$. However, if the motion of the particle is constrained, it will deviate from this applied/desired motion to satisfy the hard constraint, but this deviation will be minimum; the particle will deviate from the applied motion only by the amount that satisfies the constraint. Nature will not overdo it.

Several points are worthy of clarification here. First, in Gauss' principle, $Z$ is actually a minimum, not just stationary. Second, unlike the time-integral principle of least action, Gauss' principle is applied instantaneously (at each point in time). Third, in contrast to Hamilton's principle, Gauss' explicitly allows for non-conservative forces that do not come from a scalar work function; the impressed forces $\bm{F}_i$ can be arbitrary.

\vspace{-0.15in}
\section{Philosophy Behind Nature's Minimization of the Pressure Gradient}
\vspace{-0.15in}
Recall the Navier-Stokes equations for incompressible flows:
\begin{equation}\label{eq:N_S}
\rho\bm{a} = -\bm\nabla p + \rho\nu \Delta \bm{u}, \;\; \mbox{in}\;\Omega
\end{equation}
subject to continuity:
\begin{equation}\label{eq:Continuity}
\bm\nabla \cdot \bm{u} = 0, \;\; \mbox{in}\;\Omega
\end{equation}
where $\Omega\subset\mathbb{R}^3$ is the spatial domain, $\delta\Omega$ is its boundary, and $\bm{a}=\bm{u}_t+\bm{u}\cdot\bm\nabla \bm{u}$ is the total acceleration of the fluid particle.

\begin{figure}\vspace{-0.25in}
\includegraphics[width=8.5cm]{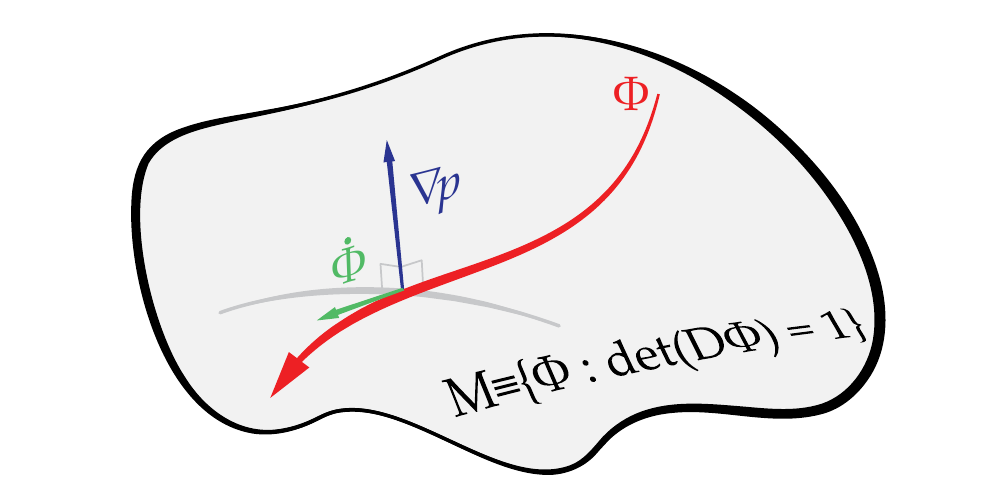}
\vspace{-0.2in}
\caption{A schematic diagram for the manifold $M$ of volume-preserving flow maps. Each point $\Phi$ on $M$ represents a flow map that takes the initial positions (Lagrangian coordinates) to their positions at some time. For an incompressible flow, these maps are volume-preserving (of unit Jacobian). A curve on this manifold represents an evolution of an incompressible flow. The tangent space is composed of divergence-free velocity vectors. The Helmholtz-Hodge decomposition implies that the pressure force $\bm\nabla p$ is orthogonal to this manifold; it is the "\textit{normal}" force that maintains the continuity constraint. Hence, by Gauss' principle, it must be minimum.}
\label{Fig:Continuity_Constraint} \vspace{-0.2in}
\end{figure}

It is noteworthy to mention that Gauss' principle is almost useless for unconstrained systems; it reduces to least-squares. Interestingly, for incompressible flows, the pressure force ($\bm\nabla p$) is a constraint force. The main role of the pressure force in incompressible flows is to enforce the continuity constraint: the divergence-free kinematic constraint on the velocity field ($\bm\nabla \cdot \bm{u} = 0$). It is straightforward to show that if $\bm{u}$ satisfies (\ref{eq:Continuity}) and the no-penetration boundary condition 
\begin{equation}\label{eq:No_Penetration_BC}
\bm{u}\cdot\bm{n}= 0, \;\; \mbox{on}\;\delta\Omega
\end{equation}
where $\bm{n}$ is the normal to the boundary, then 
\begin{equation}\label{eq:Pressure_Workless}
  \int_\Omega (\bm\nabla p \cdot \bm{u}) d\bm{x} =0,
\end{equation}
which indicates that pressure forces are workless through divergence-free velocity fields! That is, if continuity is already satisfied (the velocity field is divergence-free), the pressure forces will not affect the dynamics of this field. This fact is the main reason behind vanishing the pressure force in the first step in Chorin's standard projection method for incompressible flows \cite{Chorin_Projection}, which is based on the Helmholtz-Hodge decomposition  (e.g., \cite{Kambe2009}): a vector $\bm{v}\in\mathbb{R}^3$ can be decomposed into a divergence-free component $\bm{u}$ and a curl-free component $\bm\nabla f$ for some scalar function $f$ (i.e., $\bm{v}=\bm{u}+\bm\nabla f$). These two components are orthogonal as shown in Eq. (\ref{eq:Pressure_Workless}) provided that $\bm{u}$ satisfies the homogeneous condition (\ref{eq:No_Penetration_BC}). This decomposition can be visualized in the schematic in Fig. \ref{Fig:Continuity_Constraint}. This setup provides the basis for Arnold's seminal result \cite{Arnold_French}.

From the above discussion (and Fig. \ref{Fig:Continuity_Constraint}), it is clear that the pressure force is a constraint force. Hence, applying Gauss' principle of least constraint to the dynamics of incompressible fluids, governed by the Navier-Stokes eqaution (\ref{eq:N_S}), classifying the pressure force as a constraint force and the viscous force as an impressed force, and labeling fluid parcels with their Lagrangian coordinates $\bm\zeta$, we write the action (Gauss' quantity) as 
\[ \mathcal{A} = \frac{1}{2}\int \rho_0 \left(\bm{a}-\nu\Delta\bm{u}\right)^2 d\bm\zeta, \]
where $\rho_0=\rho_0(\bm\zeta)$ is the initial density. Realizing that $\rho J = \rho_0$, where $J$ is the Jacobian of the flow map $\bm{x}=\Phi(\bm\zeta)$ \cite{Bateman_Variational_Principle}, then the action $\mathcal{A}$ is rewritten in Eulerian coordinates as
\begin{equation}\label{eq:Gauss_Quantity}
\mathcal{A} = \frac{1}{2}\int_\Omega \rho \left(\bm{u}_t+\bm{u}\cdot\bm\nabla \bm{u}-\nu\Delta\bm{u}\right)^2 d\bm{x},
\end{equation}
which must be minimum according to Gauss' principle. Note that the Newtonian viscous force $\nu\Delta\bm{u}$ can be replaced by any arbitrary forcing; and the principle will remain applicable.

It is interesting to discuss the meaning of the minimization of $\mathcal{A}$. Note that $\mathcal{A}$ is simply $\frac{1}{2 \rho }\int_\Omega \left(\bm\nabla p\right) ^2 d\bm{x}$: the integral of the norm of the pressure gradient over the field. Since the pressure force is a constraint force (enforcing the continuity constraint), the flow field will deviate from the motion dictated by the inertial $\bm{u}\cdot\bm\nabla \bm{u}$ and viscous $\nu\Delta \bm{u}$ forces only by the amount to satisfy continuity; no larger pressure gradient will be generated than that necessary to maintain continuity. Nature will not overdo it. This new principle is what we call \textit{The Principle of Minimum Pressure Gradient} (PMPG).

The question then is: What is the independent (free) variable that minimizes $\mathcal{A}$? Interestingly, it is every ``\textit{free}"
variable! If a fluid mechanician parameterizes a flow field with some free parameters, then $\mathcal{A}$ is minimum with respect to these parameters whatever they are, as long as the representation/parameterization is admissible; i.e., it satisfies the kinematical constraint and boundary conditions. The magnitude of the pressure gradient must always be minimum. Otherwise, there will be a larger pressure gradient than necessary, which violates Nature's laws.  

In particular, to recover Navier-Stokes' equations from Gauss' principle, we adhere to the philosophy of the principle. The typical situation in particle mechanics is that the instantaneous configuration $\bm{q}(t)$ and velocity $\dot{\bm{q}}(t)$ are given, and the dynamical law must dictate the appropriate acceleration $\ddot{\bm{q}}(t)$ at this instant. As such, Gauss' principle asserts that $Z$ is minimum with respect to the free variable $\ddot{\bm{q}}(t)$ at each instant:
\[ \ddot{\bm{q}}(t)=\underset{\ddot{\bm{q}}}{\rm{argmin}} \;\; Z\left(\bm{q},\dot{\bm{q}},\ddot{\bm{q}}\right) \;\; \forall \; t.\]
Note that $Z$ is not minimum with respect to the total acceleration $\bm{a}$, which typically consists of \textit{direct} (or \textit{local}) accelerations $\ddot{\bm{q}}$ and centripetal accelerations (quadratic in velocities $\sum_j \sum_k C_{kj}(\bm{q})\dot{q}_j\dot{q}_k$ for some coefficients $C_{kj}$) because the latter component is not \textit{free}; it depends on $\dot{\bm{q}}$, which is given/fixed.

Analogously, in fluid mechanics, the instantaneous flow field $\bm{u}(t)$ is typically given and the dynamical law (e.g., Navier-Stokes') must dictate the appropriate \textit{local} acceleration $\bm{u}_t$ at this instant for the right (dynamically correct) evolution of the flow field. As such, the PMPG asserts that $\mathcal{A}$ in Eq. (\ref{eq:Gauss_Quantity}) is minimum with respect to the free variable $\bm{u}_t$ at each instant; the convective acceleration is not free. As such, we have
\begin{equation}\label{eq:Gauss_Minimization_NS}
\bm{u}_t(x;t)=\underset{\bm{u}_t}{\rm{argmin}} \;\; \frac{1}{2}\int_\Omega \rho \left(\bm{u}_t+\bm{u}\cdot\bm\nabla \bm{u}-\nu\Delta\bm{u}\right)^2 d\bm{x} \;\; \forall \; t.
\end{equation}
While the above discussion shows the philosophy behind choosing $\bm{u}_t$ as the independent (free) variable in minimizing $\mathcal{A}$, we have the following theorem whose proof is given in the Supplemental Material:

\noindent\textbf{Theorem} If $\bm{u}_t(.)$ is differentiable in $\Omega\subset\mathbb{R}^3$ and minimizes the functional 
\[ \mathcal{A}(\bm{u}_t(.)) = \frac{1}{2}\int_\Omega \rho \left(\bm{u}_t(\bm{x})+\bm{u}(\bm{x})\cdot\bm\nabla \bm{u}(\bm{x})-\nu\Delta\bm{u}(\bm{x})\right)^2 d\bm{x}\]
for all $t\in\mathbb{R}$ subject to the constraint
\[ \bm\nabla \cdot\bm{u}=0 \;\; \forall \; \bm{x}\in\Omega, \; t\in\mathbb{R},\]
and the Dirichlet boundary condition $\bm{u}(\bm{x},t)=\bm{g}(\bm{x},t)$ for all $\bm{x}\in\delta\Omega$, $t\in\mathbb{R}$, for some $\bm{g}$ differentiable in $t$, then $\bm{u}_t(.)$ must satisfy
\[\rho\left(\bm{u}_t+\bm{u}\cdot\bm\nabla \bm{u}\right)=-\bm\nabla p + \rho\nu\Delta\bm{u}\;\; \forall \; \bm{x}\in\Omega, \; t\in\mathbb{R}. \]
for some differentiable function $p$ on $\Omega$.

The above theorem implies that Navier-Stokes' equation represents the necessary condition for minimizing the pressure gradient. In fact, the pressure in the proof is the Lagrange multiplier that enforces the continuity constraint imposed on the local acceleration  $\bm\nabla \cdot\bm{u}_t=0$; the proof is straightforward, applying standard techniques from calculus of variations \cite{Burns_Optimal_Control_Book}.  

\vspace{-0.15in}
\section{Fluid Mechanics as a Minimization Problem}
\vspace{-0.15in}
In this section, we will apply the developed variational principle of minimum pressure gradient (PMPG) to solve a few classical problems in fluid mechanics---performing pure optimization without resorting to Navier-Stokes' equations. 

%\begin{figure}\vspace{-0.05in}
%\includegraphics[height=2.5cm]{Pipe_Flow.eps}
%\vspace{-0.3in}
%\caption{A schematic diagram for laminar flow in a pipe.}
%\label{Fig:Pipe_Flow} \vspace{-0.2in}
%\end{figure}

\textbf{A. Viscous Steady Case: Channel Flow}\\
Consider the simple laminar flow in a channel $y\in[-h,h]$. The velocity field is parameterized as $\bm{u}=(u(y),0)$, which automatically satisfies continuity. The flow dynamics must then dictate a specific shape for the free function $u(y)$. It is straightforward to determine this shape from the proposed PMPG without invoking the Navier-Stokes equations. Substituting by the velocity field into the action $\mathcal{A}$ in Eq. (\ref{eq:Gauss_Quantity}), we obtain
\begin{equation}\label{eq:Gauss_Pipe_Flow}
\mathcal{A}(u(.))=\frac{1}{2} \rho \nu^2 \int_{-h}^{h} \left(\frac{\partial^2 u(y)}{\partial y^2}\right)^2 dy,
\end{equation}
which should be minimum according to the PMPG. Therefore, the fluid mechanics problem is turned into the minimization problem: Find $u(y)$ that minimizes the functional (\ref{eq:Gauss_Pipe_Flow}) subject to a specified flow rate $Q$ (i.e., $\int_{-h}^{h} u(y) dy=Q$) and $u(-h)=0$ and $u(h)=0$. It is a standard calculus of variations problem (Euler's isoperimetric problem \cite{Burns_Optimal_Control_Book}) whose solution yields $\frac{\partial^2 u(y)}{\partial y^2}=\rm{constant}$, which after satisfying the boundary conditions and the flow rate constraint, results in the well-known quadratic velocity profile $u(y)=\frac{3Q}{4h}\left(1-\frac{y^2}{h^2}\right)$; the non-zero function with a minimum magnitude of second derivative $\left(\frac{\partial^2 u(y)}{\partial y^2}\right)^2$ is the quadratic function.

%\begin{figure}\vspace{-0.3in}
%\includegraphics[height=2.5cm]{Stokes_2nd_Pb.eps}
%\vspace{-0.3in}
%\caption{A schematic diagram for Stokes' second problem: A %harmonically-oscillating, infinitely-long plate.}
%\label{Fig:Stokes_2nd_Pb} \vspace{-0.2in}
%\end{figure}

\textbf{B. Viscous Unsteady Case: Stokes' Second Problem}\\
Recall the Stokes' second problem: the flow above a harmonically-oscillating, infinitely-long plate. The unsteady velocity field is parameterized as $\bm{u}=(\phi(t)\psi(y),0)$, which automatically satisfies continuity for any shapes of the free functions $\psi(y)$ and $\phi(t)$; their shapes are dictated by dynamical considerations (e.g., Navier-Stokes or the PMPG). Because there are no changes with $x$, we write the action $\mathcal{A}$ over a slice along the $y$-axis: 
\begin{equation}\label{eq:Gauss_Stokes_2nd_Pb}
\mathcal{A}(\psi(.),\phi(.))=\frac{1}{2} \rho \int_0^\infty \left(\psi(y)\dot\phi(t)-\nu\psi''(y)\phi(t)\right)^2 dy,
\end{equation}
which should be minimum for all $t$. With this modal representation $u(y,t)=\psi(y)\phi(t)$, the PMPG is capable of determining both the mode shape $\psi(y)$ and temporal coefficient $\phi(t)$ from the same minimization principle. For a given mode shape, the condition 
$\frac{\partial\mathcal{A}}{\partial\dot\phi}=0$ yields the differential equation: $\dot\phi(t) = \frac{c_1}{c_2} \phi(t)$ whose solution is $\phi(t)=\phi(0)e^{\gamma t}$, where $c_2=\rho\int_0^\infty\psi^2(y)dy\neq0$, $c_1=\nu\rho\int_0^\infty\psi(y)\psi''(y)dy$, and $\gamma=\frac{c_1}{c_2}$ are determined from the mode shape $\psi(y)$. 

Having determined the temporal solution $\phi(t)$ (up to a constant $\gamma$), the action $\mathcal{A}$ (\ref{eq:Gauss_Stokes_2nd_Pb}) can be rewritten as 
\begin{equation}\label{eq:Gauss_Stokes_2nd_Pb_Mode_Shape}
\mathcal{A}(\psi(.))=\frac{1}{2}\rho \phi^2(t) \int_0^\infty \left(\gamma\psi(y)-\nu\psi''(y)\right)^2 dy.
\end{equation}
Again, the fluid mechanics problem of computing the mode shape $\psi(y)$ turns, via the PMPG, into a pure minimization problem: Find $\psi(y)$ that makes the functional $\mathcal{A}$ in Eq. (\ref{eq:Gauss_Stokes_2nd_Pb_Mode_Shape}) stationary and satisfies the boundary conditions $\psi(0)=1$, $\psi(\infty)=0$. The solution is straightforward by applying standard techniques in calculus of variations; the Euler-Lagrange equation results in
\[ \psi''(y)=\frac{\gamma}{\nu}\psi(y) \]
whose solution, after substituting by the boundary conditions, is given by $\psi(y)=e^{-\sqrt{\frac{\gamma}{\nu}}y}$. As such, the velocity field is then written as 
\[ u(y,t)=\phi(0)e^{\gamma t} e^{-\sqrt{\frac{\gamma}{\nu}}y}. \]
Matching with the boundary condition $u(0,t)=Ue^{i\omega t}$ results in $\phi(0)=U$ and $\gamma=i\omega$, which yields the well-known solution of the Stokes' second problem \cite{Lamb}.

\textbf{C. Ideal Fluid Case: The Airfoil Problem}\\
Consider the potential flow over a two-dimensional object \citep[e.g., ][]{Schlichting}. It is straight forward to construct a flow field $\bm{u}(\bm{x};\Gamma)$ that is (i) divergence-free, (ii) irrotational, and (iii) satisfies the no-penetration boundary condition for any value of the circulation $\Gamma\in\mathbb{R}$. That is, Euler's equation does not possess a unique solution for this problem. The only theoretical fix available in the literature is through the so called \textit{Kutta condition}, which dictates that the flow is bounded everywhere. Therefore, when the body shape possesses some singularity (e.g., a sharp trailing edge in a conventional airfoil), the circulation is set to remove such a singularity. However, for singularity-free shapes (e.g., ellipse, circle), the Kutta condition is not applicable; and there is no theoretical model that can predict circulation and lift over these shapes. 

In contrast, even the inviscid version of the developed PMPG is capable of providing a unique solution over arbitrarily smooth shapes. Considering a steady snapshot (i.e., $\bm{a}=\bm{u}\cdot\bm\nabla \bm{u}$), we write the inviscid action $\mathcal{A}$ (which reduces to the \textit{Appellian}) as
\begin{equation}\label{eq:Airfoil_Appellian}
\mathcal{A}(\Gamma) = \frac{1}{2}\rho \int_\Omega \left[\bm{u}(\bm{x};\Gamma)\cdot\bm\nabla \bm{u}(\bm{x};\Gamma)\right]^2 d\bm{x}.
\end{equation}
And the PMPG yields the circulation over the airfoil as 
\begin{equation}\label{eq:Kutta_General}
\Gamma^* =\underset{\Gamma}{\rm{argmin}} \;\; \frac{1}{2}\rho \int_\Omega \left[\bm{u}(\bm{x};\Gamma)\cdot\bm\nabla \bm{u}(\bm{x};\Gamma)\right]^2 d\bm{x}.
\end{equation}
Equation (\ref{eq:Kutta_General}) provides a generalization of the Kutta-Zhukovsky condition that is, unlike the latter, derived from first principles. The PMPG allows, for the first time, computation of lift over smooth shapes without sharp edges where the Kutta condition fails. 

\begin{comment}
\begin{wrapfigure}{l}{0.250\textwidth}
\vspace{-0.15in}
 \begin{center}
 \includegraphics[height=3cm]{Gamma_v_D.pdf}
 \caption{Variation of the minimizing circulation, normalized by Kutta's, with the smoothness parameter $D$.}\vspace{-0.2in} 
 \label{Fig:Gamma_Star_D} 
 \end{center}
\end{wrapfigure}
\end{comment}

%\begin{comment}
\begin{figure}\vspace{-0.2in}
\includegraphics[height=3cm]{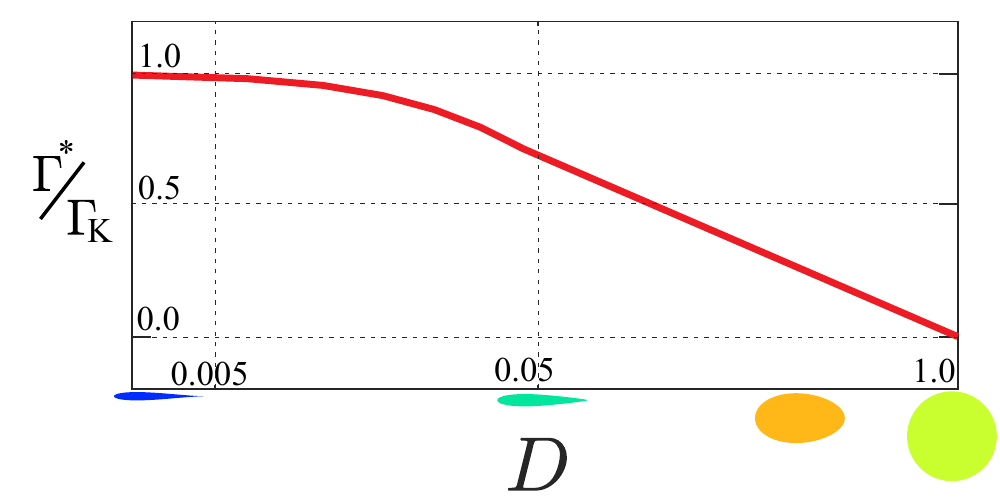}
\vspace{-0.15 in}
\caption{Variation of the minimizing circulation, normalized by Kutta's, with the smoothness parameter $D$.}\vspace{-0.2in} 
 \label{Fig:Gamma_Star_D} 
\end{figure}
%\end{comment}

\noindent Consider a family of airfoils parameterized by $D$, which controls smoothness of the trailing edge: $D=0$ results in the classical Zhukovsky airfoil with a sharp trailing edge, and $D=1$ results in a circular cylinder. Figure \ref{Fig:Gamma_Star_D} shows the variation of the minimizing circulation $\Gamma^*$ from Eq. (\ref{eq:Kutta_General}), normalized by Kutta's value $\Gamma_K$, with the parameter $D$. The figure shows that $\Gamma^*\to\Gamma_K$, as $D\to0$ (i.e., for a sharp-edged airfoil). It also shows that the PMPG recovers the classical result about the non-lifting nature of a circular cylinder in an ideal fluid: $\Gamma^*\to0$ as $D\to1$. It is remarkable that the PMPG captures the whole spectrum (from a zero lift over a circular cylinder to the Kutta-Zhukovky lift over a sharp-edged airfoil) from the same unified principle. 

The fact that the minimization principle (\ref{eq:Kutta_General}) reduces to the Kutta condition in the special case of a sharp-edged airfoil, wedded to the fact that this principle is an inviscid principle challenge the accepted wisdom about the viscous nature of the Kutta condition that prevailed over a century. In contrast, it is found that the Kutta condition is not a manifestation of viscous effects, rather of inviscid momentum effects. 

\vspace{-0.15in}
\section{Concluding Remarks}
\vspace{-0.15in}
In this paper, we developed a minimization principle of Navier-Stokes' equations that is firmly rooted in classical mechanics. The developed principle possesses an interesting physical meaning: the pressure gradient is developed only to the level that satisfies continuity; i.e., it is the least pressure gradient that maintains continuity---hence, we call it the principle of minimum pressure gradient (PMPG). The principle reduces to minimum dissipation \cite{Variational_Principles_Fluids_Existence} in the special case of ignorable inertial/convective accelerations (Stokes' flow). On the other hand, for ignorable viscous actions (Euler's dynamics), the PMPG reduces to minimum acceleration (i.e., minimum curvature). The PMPG shows how Nature balances between minimizing dissipation and curvature in the general case.

The PMPG is so generic; it is naturally written in the convenient Eulerian formulation and is applicable to 3D, unsteady, viscous flows. In fact, we proved that Navier-Stokes' equation is the necessary condition for minimizing the pressure gradient subject to the continuity constraint. Consequently, the PMPG may shed light on the Millennium Prize problem of existence of solutions of Navier-Stokes' equations; variational principles have usually been useful in studying existence of solutions of partial differential equations \cite{Functional_Analaysis}. Also, as a minimization principle, it may not suffer from a closure problem, which may smooth a path towards solving the chronic problem of turbulence closure. For example, one can easily determine the "\textit{optimal}" parameters in a RANS model by minimizing the action. In fact, one may even obtain equations for the Reynolds stress components by minimizing the pressure gradient.

We must emphasize that while we proved equivalence between the PMPG and Navier-Stokes’ equation, the former stands on its own philosophy in the light of Gauss’ principle of least constraint. Hence, the equivalence of PMPG to Navier-Stokes’ equations should not undermine the value of the principle; the equivalence between Lagrangian, Hamiltonian, and Newtonian mechanics does not imply that every formulation of a physical problem is equally tractable in each framework. The principle of least action, though equivalent to Newtonian mechanics in the ordinary scales, continues to apply to large scales (general relativity) and small scales (quantum mechanics) where Newtonian mechanics fails.  In example C (flow over an airfoil), We presented a concrete example where current formulations have failed and PMPG succeeds. In fact, the PMPG is more general than Navier-Stokes' equations for its ability to handle non-Newtonian constitutive relations and arbitrary forcing on the fluid particles.

The PMPG is expected to be of particular importance to theoretical and reduced-order modeling; a fluid mechanician can utilize his/her experience to parameterize the flow field in \textit{any} form that satisfies the kinematical constraint (continuity) and boundary conditions. Then, minimizing the action $\mathcal{A}$ with respect to the parameters of the model will provide a natural way of projecting the Navier-Stokes equations on the space of these parameters. It is interesting to recall that, after Euler developed his seminal equations that govern the dynamics of ideal fluids, Lagrange commented \cite{Dugas}
\vspace{-0.1in}
\begin{center}
``\textit{By the discovery of Euler the whole mechanics of fluids was reduced to a matter of [mathematical] analysis alone, .... Unfortunately, they are so difficult that, up to the present, it has only been possible to succeed in very special cases}”.
\end{center}
\vspace{-0.1in}
Clearly, the situation is exacerbated with Navier-Stokes' (after adding viscous forces). Interestingly, the PMPG reinstates the mechanics of fluids from pure mathematical (and computational) analysis back to the theoretical mechanics plane where the focus is not on the numerical solution of the governing equations, rather on the appropriate parameterization/representation of the flow field; it will allow fluid mechanicians to show their prowess.

%%%%%%%%%%%%%%%%%%%%%%%%%%%%%%%%%%%%%%%%%%%%%%%%%

\vspace{-0.2in}

\bibliography{aapmsamp,Aeronautical_Engineering_References,Fluid_Dynamics_References}% Produces the bibliography via BibTeX.

\begin{thebibliography}{28}%
\makeatletter
\providecommand \@ifxundefined [1]{%
 \@ifx{#1\undefined}
}%
\providecommand \@ifnum [1]{%
 \ifnum #1\expandafter \@firstoftwo
 \else \expandafter \@secondoftwo
 \fi
}%
\providecommand \@ifx [1]{%
 \ifx #1\expandafter \@firstoftwo
 \else \expandafter \@secondoftwo
 \fi
}%
\providecommand \natexlab [1]{#1}%
\providecommand \enquote  [1]{``#1''}%
\providecommand \bibnamefont  [1]{#1}%
\providecommand \bibfnamefont [1]{#1}%
\providecommand \citenamefont [1]{#1}%
\providecommand \href@noop [0]{\@secondoftwo}%
\providecommand \href [0]{\begingroup \@sanitize@url \@href}%
\providecommand \@href[1]{\@@startlink{#1}\@@href}%
\providecommand \@@href[1]{\endgroup#1\@@endlink}%
\providecommand \@sanitize@url [0]{\catcode `\\12\catcode `\$12\catcode
  `\&12\catcode `\#12\catcode `\^12\catcode `\_12\catcode `\%12\relax}%
\providecommand \@@startlink[1]{}%
\providecommand \@@endlink[0]{}%
\providecommand \url  [0]{\begingroup\@sanitize@url \@url }%
\providecommand \@url [1]{\endgroup\@href {#1}{\urlprefix }}%
\providecommand \urlprefix  [0]{URL }%
\providecommand \Eprint [0]{\href }%
\providecommand \doibase [0]{https://doi.org/}%
\providecommand \selectlanguage [0]{\@gobble}%
\providecommand \bibinfo  [0]{\@secondoftwo}%
\providecommand \bibfield  [0]{\@secondoftwo}%
\providecommand \translation [1]{[#1]}%
\providecommand \BibitemOpen [0]{}%
\providecommand \bibitemStop [0]{}%
\providecommand \bibitemNoStop [0]{.\EOS\space}%
\providecommand \EOS [0]{\spacefactor3000\relax}%
\providecommand \BibitemShut  [1]{\csname bibitem#1\endcsname}%
\let\auto@bib@innerbib\@empty
%</preamble>
\bibitem [{\citenamefont {Hargreaves}(1908)}]{Variational_Principles_Fluids1}%
  \BibitemOpen
  \bibfield  {author} {\bibinfo {author} {\bibfnamefont {R.}~\bibnamefont
  {Hargreaves}},\ }\bibfield  {title} {\bibinfo {title} {Xxxvii. a
  pressure-integral as kinetic potential},\ }\href@noop {} {\bibfield
  {journal} {\bibinfo  {journal} {The London, Edinburgh, and Dublin
  Philosophical Magazine and Journal of Science}\ }\textbf {\bibinfo {volume}
  {16}},\ \bibinfo {pages} {436} (\bibinfo {year} {1908})}\BibitemShut
  {NoStop}%
\bibitem [{\citenamefont {Serrin}(1959)}]{Variational_Principles_Fluids2}%
  \BibitemOpen
  \bibfield  {author} {\bibinfo {author} {\bibfnamefont {J.}~\bibnamefont
  {Serrin}},\ }\bibfield  {title} {\bibinfo {title} {Mathematical principles of
  classical fluid mechanics},\ }in\ \href@noop {} {\emph {\bibinfo {booktitle}
  {Fluid Dynamics I/Str{\"o}mungsmechanik I}}}\ (\bibinfo  {publisher}
  {Springer},\ \bibinfo {year} {1959})\ pp.\ \bibinfo {pages}
  {125--263}\BibitemShut {NoStop}%
\bibitem [{\citenamefont
  {Penfield~Jr}(1966)}]{Variational_Principles_Fluids_Hamilton_PoF}%
  \BibitemOpen
  \bibfield  {author} {\bibinfo {author} {\bibfnamefont {P.}~\bibnamefont
  {Penfield~Jr}},\ }\bibfield  {title} {\bibinfo {title} {Hamilton's principle
  for fluids},\ }\href@noop {} {\bibfield  {journal} {\bibinfo  {journal} {The
  Physics of Fluids}\ }\textbf {\bibinfo {volume} {9}},\ \bibinfo {pages}
  {1184} (\bibinfo {year} {1966})}\BibitemShut {NoStop}%
\bibitem [{\citenamefont {Seliger}\ and\ \citenamefont
  {Whitham}(1968)}]{Seliger_Variational_Principles}%
  \BibitemOpen
  \bibfield  {author} {\bibinfo {author} {\bibfnamefont {R.~L.}\ \bibnamefont
  {Seliger}}\ and\ \bibinfo {author} {\bibfnamefont {G.~B.}\ \bibnamefont
  {Whitham}},\ }\bibfield  {title} {\bibinfo {title} {Variational principles in
  continuum mechanics},\ }in\ \href@noop {} {\emph {\bibinfo {booktitle}
  {Proceedings of the Royal Society of London A: Mathematical, Physical and
  Engineering Sciences}}},\ Vol.\ \bibinfo {volume} {305}\ (\bibinfo
  {organization} {The Royal Society},\ \bibinfo {year} {1968})\ pp.\ \bibinfo
  {pages} {1--25}\BibitemShut {NoStop}%
\bibitem [{\citenamefont {Bretherton}(1970)}]{Variational_Principles_Fluids5}%
  \BibitemOpen
  \bibfield  {author} {\bibinfo {author} {\bibfnamefont {F.~P.}\ \bibnamefont
  {Bretherton}},\ }\bibfield  {title} {\bibinfo {title} {A note on hamilton's
  principle for perfect fluids},\ }\href@noop {} {\bibfield  {journal}
  {\bibinfo  {journal} {Journal of Fluid Mechanics}\ }\textbf {\bibinfo
  {volume} {44}},\ \bibinfo {pages} {19} (\bibinfo {year} {1970})}\BibitemShut
  {NoStop}%
\bibitem [{\citenamefont {Salmon}(1988)}]{Hamiltonian_Fluid_Mechanics}%
  \BibitemOpen
  \bibfield  {author} {\bibinfo {author} {\bibfnamefont {R.}~\bibnamefont
  {Salmon}},\ }\bibfield  {title} {\bibinfo {title} {Hamiltonian fluid
  mechanics},\ }\href@noop {} {\bibfield  {journal} {\bibinfo  {journal}
  {Annual review of fluid mechanics}\ }\textbf {\bibinfo {volume} {20}},\
  \bibinfo {pages} {225} (\bibinfo {year} {1988})}\BibitemShut {NoStop}%
\bibitem [{\citenamefont
  {Morrison}(1998)}]{Variational_Principles_Morrison_Review}%
  \BibitemOpen
  \bibfield  {author} {\bibinfo {author} {\bibfnamefont {P.~J.}\ \bibnamefont
  {Morrison}},\ }\bibfield  {title} {\bibinfo {title} {Hamiltonian description
  of the ideal fluid},\ }\href@noop {} {\bibfield  {journal} {\bibinfo
  {journal} {Reviews of modern physics}\ }\textbf {\bibinfo {volume} {70}},\
  \bibinfo {pages} {467} (\bibinfo {year} {1998})}\BibitemShut {NoStop}%
\bibitem [{\citenamefont
  {Yasue}(1983)}]{Variational_Principles_Fluids_Stochastic}%
  \BibitemOpen
  \bibfield  {author} {\bibinfo {author} {\bibfnamefont {K.}~\bibnamefont
  {Yasue}},\ }\bibfield  {title} {\bibinfo {title} {A variational principle for
  the navier-stokes equation},\ }\href@noop {} {\bibfield  {journal} {\bibinfo
  {journal} {Journal of Functional Analysis}\ }\textbf {\bibinfo {volume}
  {51}},\ \bibinfo {pages} {133} (\bibinfo {year} {1983})}\BibitemShut
  {NoStop}%
\bibitem [{\citenamefont {Kerswell}(1999)}]{Variational_Principles_NS}%
  \BibitemOpen
  \bibfield  {author} {\bibinfo {author} {\bibfnamefont {R.~R.}\ \bibnamefont
  {Kerswell}},\ }\bibfield  {title} {\bibinfo {title} {Variational principle
  for the navier-stokes equations},\ }\href@noop {} {\bibfield  {journal}
  {\bibinfo  {journal} {Physical Review E}\ }\textbf {\bibinfo {volume} {59}},\
  \bibinfo {pages} {5482} (\bibinfo {year} {1999})}\BibitemShut {NoStop}%
\bibitem [{\citenamefont
  {Gomes}(2005)}]{Variational_Principles_Fluids_Stochastic_Gomes}%
  \BibitemOpen
  \bibfield  {author} {\bibinfo {author} {\bibfnamefont {D.~A.}\ \bibnamefont
  {Gomes}},\ }\bibfield  {title} {\bibinfo {title} {A variational formulation
  for the navier-stokes equation},\ }\href@noop {} {\bibfield  {journal}
  {\bibinfo  {journal} {Communications in mathematical physics}\ }\textbf
  {\bibinfo {volume} {257}},\ \bibinfo {pages} {227} (\bibinfo {year}
  {2005})}\BibitemShut {NoStop}%
\bibitem [{\citenamefont
  {Eyink}(2010)}]{Variational_Principles_Fluids_Stochastic2}%
  \BibitemOpen
  \bibfield  {author} {\bibinfo {author} {\bibfnamefont {G.~L.}\ \bibnamefont
  {Eyink}},\ }\bibfield  {title} {\bibinfo {title} {Stochastic least-action
  principle for the incompressible navier--stokes equation},\ }\href@noop {}
  {\bibfield  {journal} {\bibinfo  {journal} {Physica D: Nonlinear Phenomena}\
  }\textbf {\bibinfo {volume} {239}},\ \bibinfo {pages} {1236} (\bibinfo {year}
  {2010})}\BibitemShut {NoStop}%
\bibitem [{\citenamefont {Fukagawa}\ and\ \citenamefont
  {Fujitani}(2012)}]{Variational_Principles_NS_Nonholonomic}%
  \BibitemOpen
  \bibfield  {author} {\bibinfo {author} {\bibfnamefont {H.}~\bibnamefont
  {Fukagawa}}\ and\ \bibinfo {author} {\bibfnamefont {Y.}~\bibnamefont
  {Fujitani}},\ }\bibfield  {title} {\bibinfo {title} {A variational principle
  for dissipative fluid dynamics},\ }\href@noop {} {\bibfield  {journal}
  {\bibinfo  {journal} {Progress of Theoretical Physics}\ }\textbf {\bibinfo
  {volume} {127}},\ \bibinfo {pages} {921} (\bibinfo {year}
  {2012})}\BibitemShut {NoStop}%
\bibitem [{\citenamefont {Galley}\ \emph {et~al.}(2014)\citenamefont {Galley},
  \citenamefont {Tsang},\ and\ \citenamefont
  {Stein}}]{Variational_Principles_NS_2n}%
  \BibitemOpen
  \bibfield  {author} {\bibinfo {author} {\bibfnamefont {C.~R.}\ \bibnamefont
  {Galley}}, \bibinfo {author} {\bibfnamefont {D.}~\bibnamefont {Tsang}},\ and\
  \bibinfo {author} {\bibfnamefont {L.~C.}\ \bibnamefont {Stein}},\ }\bibfield
  {title} {\bibinfo {title} {The principle of stationary nonconservative action
  for classical mechanics and field theories},\ }\href@noop {} {\bibfield
  {journal} {\bibinfo  {journal} {arXiv preprint arXiv:1412.3082}\ } (\bibinfo
  {year} {2014})}\BibitemShut {NoStop}%
\bibitem [{\citenamefont {Gay-Balmaz}\ and\ \citenamefont
  {Yoshimura}(2017)}]{Variational_Principles_NS_Nonholonomic2}%
  \BibitemOpen
  \bibfield  {author} {\bibinfo {author} {\bibfnamefont {F.}~\bibnamefont
  {Gay-Balmaz}}\ and\ \bibinfo {author} {\bibfnamefont {H.}~\bibnamefont
  {Yoshimura}},\ }\bibfield  {title} {\bibinfo {title} {A lagrangian
  variational formulation for nonequilibrium thermodynamics. part ii: continuum
  systems},\ }\href@noop {} {\bibfield  {journal} {\bibinfo  {journal} {Journal
  of Geometry and Physics}\ }\textbf {\bibinfo {volume} {111}},\ \bibinfo
  {pages} {194} (\bibinfo {year} {2017})}\BibitemShut {NoStop}%
\bibitem [{\citenamefont {Papastavridis}(2014)}]{Papastavridis2014}%
  \BibitemOpen
  \bibfield  {author} {\bibinfo {author} {\bibfnamefont {J.}~\bibnamefont
  {Papastavridis}},\ }\href@noop {} {\emph {\bibinfo {title} {{Analytical
  mechanics: a comprehensive treatise on the dynamics of constrained systems --
  Reprint edition.}}}}\ (\bibinfo  {publisher} {Word Scientific Publishing
  Company},\ \bibinfo {year} {2014})\BibitemShut {NoStop}%
\bibitem [{\citenamefont {Lilov}\ and\ \citenamefont
  {Lorer}(1982)}]{Gauss_21century}%
  \BibitemOpen
  \bibfield  {author} {\bibinfo {author} {\bibfnamefont {L.}~\bibnamefont
  {Lilov}}\ and\ \bibinfo {author} {\bibfnamefont {M.}~\bibnamefont {Lorer}},\
  }\bibfield  {title} {\bibinfo {title} {Dynamic analysis of multirigid-body
  system based on the gauss principle},\ }\href@noop {} {\bibfield  {journal}
  {\bibinfo  {journal} {ZAMM-Journal of Applied Mathematics and
  Mechanics/Zeitschrift f{\"u}r Angewandte Mathematik und Mechanik}\ }\textbf
  {\bibinfo {volume} {62}},\ \bibinfo {pages} {539} (\bibinfo {year}
  {1982})}\BibitemShut {NoStop}%
\bibitem [{\citenamefont {Udwadia}\ and\ \citenamefont
  {Kalaba}(1996)}]{Udwadia_Kalaba_Book}%
  \BibitemOpen
  \bibfield  {author} {\bibinfo {author} {\bibfnamefont {F.~E.}\ \bibnamefont
  {Udwadia}}\ and\ \bibinfo {author} {\bibfnamefont {R.~E.}\ \bibnamefont
  {Kalaba}},\ }\href@noop {} {\emph {\bibinfo {title} {Analytical dynamics}}}\
  (\bibinfo {year} {1996})\BibitemShut {NoStop}%
\bibitem [{\citenamefont {Lanczos}(1970)}]{Lanczos_Variational_Mechanics_Book}%
  \BibitemOpen
  \bibfield  {author} {\bibinfo {author} {\bibfnamefont {C.}~\bibnamefont
  {Lanczos}},\ }\href@noop {} {\emph {\bibinfo {title} {{The variational
  principles of mechanics}}}}\ (\bibinfo  {publisher} {Courier Corporation},\
  \bibinfo {year} {1970})\BibitemShut {NoStop}%
\bibitem [{\citenamefont {Chorin}(1968)}]{Chorin_Projection}%
  \BibitemOpen
  \bibfield  {author} {\bibinfo {author} {\bibfnamefont {A.~J.}\ \bibnamefont
  {Chorin}},\ }\bibfield  {title} {\bibinfo {title} {Numerical solution of the
  navier-stokes equations},\ }\href@noop {} {\bibfield  {journal} {\bibinfo
  {journal} {Mathematics of computation}\ }\textbf {\bibinfo {volume} {22}},\
  \bibinfo {pages} {745} (\bibinfo {year} {1968})}\BibitemShut {NoStop}%
\bibitem [{\citenamefont {Kambe}(2009)}]{Kambe2009}%
  \BibitemOpen
  \bibfield  {author} {\bibinfo {author} {\bibfnamefont {T.}~\bibnamefont
  {Kambe}},\ }\href@noop {} {\emph {\bibinfo {title} {{Geometrical theory of
  dynamical systems and fluid flows}}}},\ Vol.~\bibinfo {volume} {23}\
  (\bibinfo  {publisher} {World Scientific Publishing Co Inc},\ \bibinfo {year}
  {2009})\BibitemShut {NoStop}%
\bibitem [{\citenamefont {Arnold}(1966)}]{Arnold_French}%
  \BibitemOpen
  \bibfield  {author} {\bibinfo {author} {\bibfnamefont {V.~I.}\ \bibnamefont
  {Arnold}},\ }\bibfield  {title} {\bibinfo {title} {Sur la g{\'e}om{\'e}trie
  diff{\'e}rentielle des groupes de lie de dimension infinie et ses
  applications {\`a} l'hydrodynamique des fluides parfaits},\ }in\ \href@noop
  {} {\emph {\bibinfo {booktitle} {Annales de l'institut Fourier}}},\
  Vol.~\bibinfo {volume} {16}\ (\bibinfo {year} {1966})\ pp.\ \bibinfo {pages}
  {319--361}\BibitemShut {NoStop}%
\bibitem [{\citenamefont {Bateman}(1929)}]{Bateman_Variational_Principle}%
  \BibitemOpen
  \bibfield  {author} {\bibinfo {author} {\bibfnamefont {H.}~\bibnamefont
  {Bateman}},\ }\bibfield  {title} {\bibinfo {title} {Notes on a differential
  equation which occurs in the two-dimensional motion of a compressible fluid
  and the associated variational problems},\ }\href@noop {} {\bibfield
  {journal} {\bibinfo  {journal} {Proc. R. Soc. Lond. A}\ }\textbf {\bibinfo
  {volume} {125}},\ \bibinfo {pages} {598} (\bibinfo {year}
  {1929})}\BibitemShut {NoStop}%
\bibitem [{\citenamefont {Burns}(2013)}]{Burns_Optimal_Control_Book}%
  \BibitemOpen
  \bibfield  {author} {\bibinfo {author} {\bibfnamefont {J.~A.}\ \bibnamefont
  {Burns}},\ }\href@noop {} {\emph {\bibinfo {title} {Introduction to the
  calculus of variations and control with modern applications}}}\ (\bibinfo
  {publisher} {CRC Press},\ \bibinfo {year} {2013})\BibitemShut {NoStop}%
\bibitem [{\citenamefont {Lamb}(1932)}]{Lamb}%
  \BibitemOpen
  \bibfield  {author} {\bibinfo {author} {\bibfnamefont {H.}~\bibnamefont
  {Lamb}},\ }\href@noop {} {\emph {\bibinfo {title} {Hydrodynamics}}}\
  (\bibinfo  {publisher} {Cambridge university press},\ \bibinfo {year}
  {1932})\BibitemShut {NoStop}%
\bibitem [{\citenamefont {Schlichting}\ and\ \citenamefont
  {Truckenbrodt}(1979)}]{Schlichting}%
  \BibitemOpen
  \bibfield  {author} {\bibinfo {author} {\bibfnamefont {H.}~\bibnamefont
  {Schlichting}}\ and\ \bibinfo {author} {\bibfnamefont {E.}~\bibnamefont
  {Truckenbrodt}},\ }\href@noop {} {\emph {\bibinfo {title} {Aerodynamics of
  the Airplane}}}\ (\bibinfo  {publisher} {McGraw-Hill},\ \bibinfo {year}
  {1979})\BibitemShut {NoStop}%
\bibitem [{\citenamefont
  {Finlayson}(1972)}]{Variational_Principles_Fluids_Existence}%
  \BibitemOpen
  \bibfield  {author} {\bibinfo {author} {\bibfnamefont {B.~A.}\ \bibnamefont
  {Finlayson}},\ }\bibfield  {title} {\bibinfo {title} {Existence of
  variational principles for the navier-stokes equation},\ }\href@noop {}
  {\bibfield  {journal} {\bibinfo  {journal} {The physics of fluids}\ }\textbf
  {\bibinfo {volume} {15}},\ \bibinfo {pages} {963} (\bibinfo {year}
  {1972})}\BibitemShut {NoStop}%
\bibitem [{\citenamefont {Reed}\ and\ \citenamefont
  {Simon}(1980)}]{Functional_Analaysis}%
  \BibitemOpen
  \bibfield  {author} {\bibinfo {author} {\bibfnamefont {M.}~\bibnamefont
  {Reed}}\ and\ \bibinfo {author} {\bibfnamefont {B.}~\bibnamefont {Simon}},\
  }\href@noop {} {\emph {\bibinfo {title} {Methods of modern mathematical
  physics. vol. 1. Functional analysis}}}\ (\bibinfo  {publisher} {Academic New
  York},\ \bibinfo {year} {1980})\BibitemShut {NoStop}%
\bibitem [{\citenamefont {Dugas}(1988)}]{Dugas}%
  \BibitemOpen
  \bibfield  {author} {\bibinfo {author} {\bibfnamefont {R.}~\bibnamefont
  {Dugas}},\ }\href@noop {} {\emph {\bibinfo {title} {A History of Mechanics,
  translated into English by JR Maddox. NY}}}\ (\bibinfo  {publisher} {Dover
  Publications, Inc},\ \bibinfo {year} {1988})\BibitemShut {NoStop}%
\end{thebibliography}%

\end{document}